\documentclass[aps,prl,showkeys,floatfix,twocolumn,bibnotes,nofootinbib,superscriptaddress]{revtex4-2}
\usepackage{graphicx,amsmath,amssymb,hyperref,natbib,slashed}
\usepackage[dvipsnames]{xcolor}
\usepackage{xspace}

\usepackage[utf8]{inputenc}

\definecolor{burgundy}{rgb}{0.5, 0.0, 0.13}

\newcommand{\eVdist}{\kern-0.06em}

\newcommand{\D}{\mathrm{d}}

\newcommand{\be}{\begin{equation}}
\newcommand{\ee}{\end{equation}}
\newcommand{\bea}{\begin{eqnarray}}
\newcommand{\eea}{\end{eqnarray}}

\begin{document}
\hfill DESY 21-044, ULB-TH/21-03

\title{Dark Matter from Exponential Growth\\[0.2cm]\small{Originally proposed as Pandemic Dark Matter}}

\newcommand{\AddrOslo}{%
Department of Physics, University of Oslo, Box 1048, N-0316 Oslo, Norway}
\newcommand{\AddrDESY}{%
Deutsches Elektronen-Synchrotron DESY,  Notkestra\ss e 85, D-22607 Hamburg, Germany}
\newcommand{\AddrBrussels}{%
Service de Physique Théorique, Université Libre de Bruxelles, Boulevard du Triomphe, CP225, B-1050 Brussels, Belgium}
\newcommand{\AddrNYU}{%
Center for Cosmology and Particle Physics, Department of Physics, New York University, New York, NY 10003, USA}
\newcommand{\AddrKITP}{%
Kavli Institute for Theoretical Physics, University of California, Santa Barbara, CA 93106, USA}
\newcommand{\AddTAU}{%
School of Physics and Astronomy, Tel-Aviv University, Tel-Aviv 69978, Israel
\vspace*{0.1cm}}

 \author{Torsten Bringmann}
 \email{torsten.bringmann@fys.uio.no}
\affiliation{\AddrOslo}

  \author{Paul Frederik Depta}
 \email{frederik.depta@desy.de}
 \affiliation{\AddrDESY}

  \author{Marco Hufnagel}
 \email{marco.hufnagel@ulb.ac.be}
\affiliation{\AddrBrussels}

  \author{Joshua T.\ Ruderman}
 \email{ruderman@nyu.edu}
\affiliation{\AddrNYU}
\affiliation{\AddrDESY}
\affiliation{\AddrKITP}
\affiliation{\AddTAU}

 \author{Kai Schmidt-Hoberg}
 \email{kai.schmidt-hoberg@desy.de}
\affiliation{\AddrDESY}

\begin{abstract}
We propose a novel mechanism for the
production of dark matter (DM) from a thermal bath, based on the idea that DM particles $\chi$ 
can transform 
heat bath particles $\psi$: $\chi \psi \rightarrow \chi \chi$.  For a small initial abundance of $\chi$ 
this leads to an exponential growth of the DM number density, in close analogy to
other familiar exponential growth processes in nature.
We demonstrate that this mechanism complements freeze-in and freeze-out production in a generic way, 
opening new parameter space to explain the observed DM abundance,
and we discuss observational prospects for such scenarios.
\end{abstract}


\maketitle

\paragraph*{Introduction.---}%

While the identity and underlying properties of the dark matter (DM) in our Universe remain mysterious, 
its energy density has been precisely inferred by a series of satellite missions studying the Cosmic Microwave Background (CMB).
Any theoretical description of DM must therefore include a \textit{DM production mechanism}
which leads to the observed DM relic abundance of $\Omega_{\text{DM}} h^2 \simeq 0.12$~\cite{Aghanim:2018eyx}.

A particularly appealing framework for the genesis of DM, minimizing the dependence on initial conditions, 
is its creation out of a thermal bath. 
The most commonly adopted paradigm falling into this category is thermal freeze-out from the 
primordial plasma of Standard Model (SM) particles in the early Universe~\cite{Lee:1977ua}.
However, given the increasingly strong constraints on this setup, 
a plethora of alternate production scenarios with DM initially in thermal equilibrium have recently been proposed,
including 
\textit{hidden sector freeze-out}~\cite{Finkbeiner:2007kk,Pospelov:2007mp,Feng:2008mu,Pospelov:2008zw,Sigurdson:2009uz,Cheung:2010gj,Bringmann:2020mgx}, 
\textit{Forbidden DM}~\cite{DAgnolo:2015ujb,DAgnolo:2020mpt},  \textit{Cannibal DM}~\cite{Pappadopulo:2016pkp,Farina:2016llk}, 
\textit{Coscattering DM}~\cite{DAgnolo:2017dbv,Garny:2017rxs,DAgnolo:2019zkf}, \textit{Zombie DM}~\cite{Kramer:2021hal}, 
\textit{Elder DM}~\cite{Kuflik:2015isi}, \textit{Kinder DM}~\cite{Fitzpatrick:2020vba},
and \textit{SIMP DM}~\cite{Carlson:1992fn,Hochberg:2014dra,Smirnov:2020zwf}.
Another possibility is that DM never entered thermal equilibrium at all, 
in which case it can be produced via a `leakage' out of a thermal bath, often referred to as 
\textit{freeze-in}~\cite{Hall:2009bx,Chu:2011be}. While a large number of variants of the freeze-out paradigm have 
been suggested, less model building has been performed around the freeze-in idea (see however 
Refs.~\cite{Chu:2013jja,Falkowski:2017uya,An:2018nvz,Hambye:2019dwd,Mondino:2020lsc,Belanger:2020npe,Bernal:2020gzm,March-Russell:2020nun}).

In this letter we propose a novel and generic
DM production scenario between these two polarities, based on the idea that 
a DM particle $\chi$ can `transform' a heat bath particle $\psi$ into 
another $\chi$, cf.\ Fig.~\ref{fig:infection}.
For a small initial abundance $n_\chi$, as shown below, this results in an exponential growth of the DM abundance. 
To be in accord with observations the exponential growth is required to 
shut off before the DM particle $\chi$ is fully thermalized,
so one can also think of this mechanism as a `failed thermalization'. Interestingly, the exponential growth of $n_\chi$ 
comes to an end naturally in our framework, 
so that the observed DM abundance is readily obtained.

\begin{figure}[t]
\includegraphics[width=0.6\columnwidth]{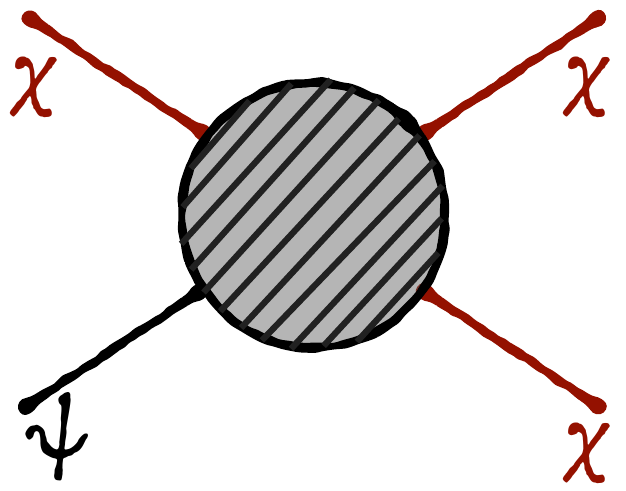}
\caption{%
The 
transformation process leading to exponential production of DM ($\chi$) from the heat bath
($\psi$).
}
\label{fig:infection}
\end{figure}

\smallskip
\paragraph*{A novel DM production mechanism.---}%
Quantitatively, the evolution of the DM number 
density $n_\chi$ is governed by the Boltzmann equation
\be
\label{eq:boltz1}
\dot n_\chi+3Hn_\chi = \langle\sigma v\rangle_\mathrm{tr}n_\chi n_\psi^\mathrm{eq}-
\langle\sigma v\rangle_{\mathrm{an}}n_\chi^2\,.
\ee
Here, $n_{\psi}^\mathrm{eq}$ is the number density of $\psi$ in equilibrium,  $H$ is the Hubble rate, and 
$\langle\sigma v\rangle_\mathrm{tr}$ ($\langle\sigma v\rangle_\mathrm{an}$) is the 
cross-section for the process $\chi\psi\to\chi\chi$ ($\chi\chi\to\chi\psi$), averaged over the
phase-space of the initial state.  We assume that $\psi$ is in equilibrium with the SM heat bath; this 
equilibrium can be maintained for example by rapid annihilations of $\psi \psi$ to SM states.
We note that the zombie collisions of Refs.~\cite{Kramer:2021hal,Berlin:2017ife} involve a  similar 
process to Fig.~\ref{fig:infection}, but the roles of the DM and bath particles are reversed.

As long as $n_\chi^\mathrm{eq}\gg n_\chi$, we can neglect the second term on the r.h.s.~of
the above equation.  
Introducing $x_\psi\equiv m_\psi/T$ and $Y_\chi\equiv n_\chi/s$, with 
$s$ being the entropy density of the heat bath, 
the solution of the Boltzmann equation is given by
\be
\label{eq:transmission_evolution}
 Y_\chi(x_\psi) \simeq Y_\chi^0\exp\left[\int_{x^0_\psi}^{x_\psi}\frac{\D x}{x} P(x)  \right]\,,
\ee
where
\be \label{eq:P}
P(x)={\tilde H}^{-1}\, {n^\mathrm{eq}_\psi}  \langle\sigma v\rangle_\mathrm{tr}\,.
\ee
Here
$Y^0_\chi$ denotes the DM abundance at some initial
`time' $x^0_\psi$, and we have defined $\tilde H\equiv H/\left[1+ (1/3)d(\log g^s_{\rm eff})/d(\log T)\right]$, 
where $g^s_{\rm eff}$ encodes the entropy degrees of freedom. 

Eq.~(\ref{eq:transmission_evolution}) describes exponential growth of the DM abundance,
with growth rate $P$, as long as $P'(x)>0$. 
For highly relativistic heat bath particles (with $n_\psi^\mathrm{eq}\propto x_\psi^{-3}$)
this is automatically achieved for  $\langle\sigma v\rangle_\mathrm{tr}=(\sigma v)^0_\mathrm{tr}\, x^k_\psi$
with $(\sigma v)^0_\mathrm{tr}\simeq const.$ and $k>1$, i.e.~infrared (IR) dominated 
transformation processes 
since $H\propto x_\psi^{-2}$. Later,
once the heat bath particles become non-relativistic, exponential growth will inevitably
come to an end for any value of $k$ due to the Boltzmann suppression of $n_\psi^{\rm eq}$,
leading to an asymptotically flat $Y_\chi(x_\psi)$. Parametrically, we thus find 
\be
Y_\chi^\mathrm{final}\sim Y_\chi^0 \exp\left[\frac{\lambda_*}{k\!-\!1} x_{\psi,\mathrm{NR}}^{k-1}\right]
\label{eq:Yanalytic}
\ee
for the final DM abundance, where $x_{\psi,\mathrm{NR}}\sim3$ 
and $\lambda_*\sim 6\times10^{-2} g_\psi g_{\rm eff}^{-\frac12}m_\psi m_\mathrm{Pl} (\sigma v)^0_\mathrm{tr}$,
 with $g_{\rm eff}$ the energy degrees of freedom and
$m_\mathrm{Pl}$ the Planck mass.

\begin{figure}[t!]
\includegraphics[width=0.95\columnwidth]{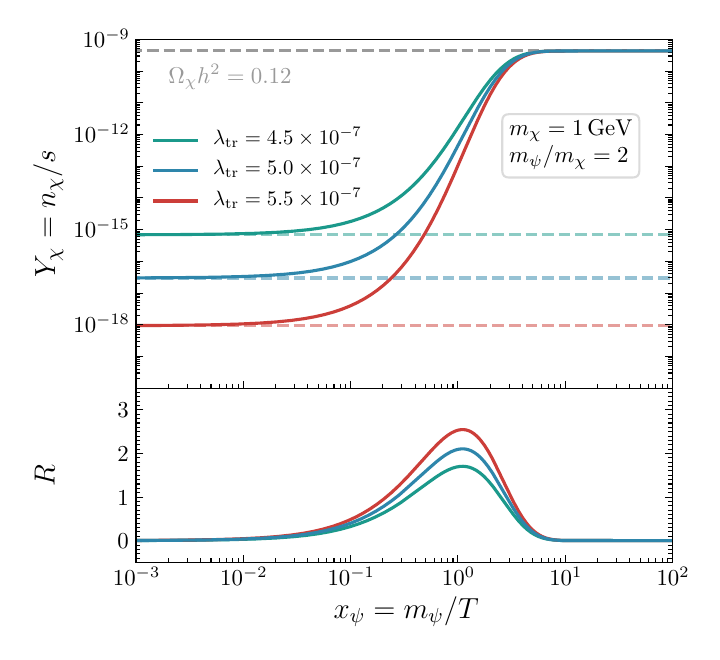}
\caption{%
{\it Top.} Number density of $\chi$ relative to the entropy density of the heat bath (solid lines) for 
$m_\chi=1$\,GeV, $m_\psi=2$\,GeV\@, and different values of the transformation 
coupling $\lambda_\text{tr}$. 
For each value of $\lambda_\text{tr}$, we fix the initial abundance 
of $\chi$ (dashed lines) such that the final abundance corresponds to the observed DM density.
{\it Bottom.} $R$-value corresponding to the 
abundance evolution
in the top panel. 
}
\label{fig:epi_simp}
\end{figure}

We confirm this expectation in Fig.~\ref{fig:epi_simp} where we show
(with solid lines) the full solution of Eq.~(\ref{eq:boltz1}), adopting for illustration
a constant
amplitude 
$\left|\mathcal{M}_\mathrm{tr}\right|^2=\lambda_\mathrm{tr}^2$, which is realized if $\chi$ and $\psi$ 
are real scalars with interaction $\mathcal{L}\supset(\lambda_\mathrm{tr}/3!) \psi \chi^3$.
We calculate, for simplicity,  $\langle\sigma v\rangle_\mathrm{tr}$ for a Maxwell-Boltzmann 
distribution~\cite{Edsjo:1997bg},
leaving a determination of its precise phase space distribution for future work; 
this leads to $\langle\sigma v\rangle_\mathrm{tr}\propto T^{-2}$ for $T\gg m_\chi,m_\psi$.
Starting from an initial value of the DM abundance (indicated by the dashed lines),
the onset of exponential growth is clearly visible as $P$ becomes larger than $\sim0.2$
for $x_\psi\gtrsim0.01$ -- until it stalls because $P$ is heavily suppressed again for 
$x_\psi\gtrsim5$.  The figure also illustrates an attractive feature of exponential 
growth from a phenomenological point of view: the coupling strength 
required to match the observed DM relic abundance is only \textit{logarithmically} sensitive to the
initial abundance. In the examples shown here, e.g., decreasing the initial abundance by four
orders of magnitude (from the green to the red line) is compensated by a mere increase of 
about 22\,\% in $\lambda_\mathrm{tr}$.

\smallskip
\paragraph*{Exponential growth in nature.---}%
It is 
intriguing how closely the evolution of the DM abundance in Fig.~\ref{fig:epi_simp} 
mimics other well-known examples of exponential growth in nature, like
for example the progression of an illness after an initial outbreak.
In fact, we can formalize this analogy by referring to the SIR epidemiological 
model~\cite{KermackMcKendrick},
where the number of infected individuals, $I$, evolves according to
\be \label{eq:SIR}
\dot I = \beta S I - \gamma I \, ,
\ee
with $S$ the number of susceptible individuals and $\beta$ and $\gamma$ the infection and 
recovery rates, respectively.  
We recognize that this is simply  Eq.~\eqref{eq:boltz1}, in the limit $n_\chi \ll n_\chi^{\rm eq}$,
after identifying: $I \leftrightarrow n_\chi$, $S \leftrightarrow n_\psi^{\rm eq}$, 
$\beta \leftrightarrow \langle\sigma v\rangle_\mathrm{tr}$, and $\gamma \leftrightarrow 3H$.  
This 
mathematically exact correspondence motivates us to further introduce 
\be \label{eq:R}
R \equiv \frac{\beta S}{\gamma} = \frac{n_\psi \langle\sigma v\rangle_\mathrm{tr}}{3 H} = \frac{\tilde H}{3H} P \, \, ,
\ee
where the final equality follows from Eq.~\eqref{eq:P}.
$R$ measures the number of transformation processes that each DM particle undergoes per Hubble time
and is,  through Eqs.~\eqref{eq:transmission_evolution} and \eqref{eq:R}, 
 directly related to the final DM abundance.

\smallskip
\paragraph*{Initial abundance.---}%
In the above discussion we have deliberately remained agnostic about the origin of $Y_\chi^0$, and
simply treated this quantity as a free input parameter. We now outline various physical mechanisms 
that could generate such an initial DM abundance.

The first class of initial 
DM production mechanisms takes place 
much earlier than the typically rather short period where 
$\chi\psi\to\chi\chi$ processes dominate.
This includes well-studied examples such as UV-dominated freeze-in~\cite{Moroi:1993mb,Bolz:2000fu} or 
direct production from the decay of the inflaton or other heavy particles~\cite{Takahashi:2007tz} 
 -- but could also be related to more 
exotic examples like 
false vacua after a phase transition 
in the dark sector~\cite{Witten:1984rs,Asadi:2021pwo} 
or by the evaporation of black holes. 
Common to all these scenarios is that the final DM abundance is independent of how exactly the initial 
abundance is set:
the only phenomenologically relevant input is the DM abundance at the onset of the
era of exponential growth, thus providing a direct map to the generic situation depicted in Fig.~\ref{fig:epi_simp}.

\begin{figure}[t!]
\includegraphics[width=0.95\columnwidth]{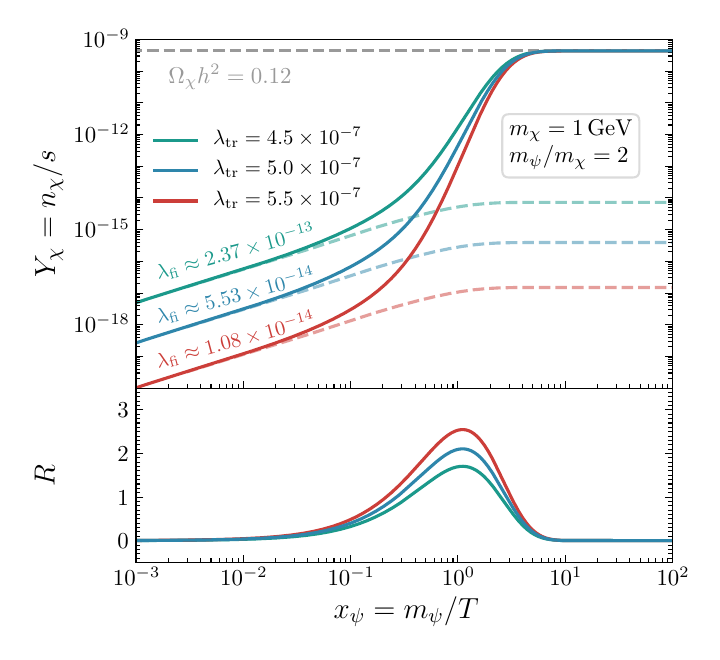}
\caption{%
As in Fig.~\ref{fig:epi_simp}, but now with a vanishing initial DM abundance
and, on top of the transformation 
interaction, freeze-in production based on a
constant matrix element. The coupling $\lambda_{\rm fi}$ for the latter is chosen such that the final 
abundance of $\chi$ corresponds to 
the observed DM density. 
Dashed lines show the would-be abundance from freeze-in alone
(when setting $\lambda_{\rm tr}=0$, for which $\lambda_\text{fi} \approx 5.81\times 10^{-11}$
would give $\Omega_{\text{DM}} h^2 = 0.12$).
}
\label{fig:epi_simp_fi}
\end{figure}

In the second class of relevant scenarios, 
the initial and exponential phases of DM production are 
intertwined. This is particularly relevant for IR dominated 
freeze-in rates~\cite{Hall:2009bx} 
that are too small to explain the observed DM abundance without
a subsequent phase of exponential growth. 
A nice feature of the 
mechanism proposed here is in fact that a non-vanishing freeze-in contribution
due to the transformation 
coupling $\lambda_\mathrm{tr}$
is automatically built-in, as discussed below.
In general, the Boltzmann equation including $2\to2$
freeze-in processes becomes
\be
\label{eq:boltz2}
  \dot n_\chi+3Hn_\chi \simeq \langle\sigma v\rangle_\mathrm{tr}\,n_\chi n_\psi^\mathrm{eq} 
  +  \langle\sigma v\rangle_\mathrm{fi}\,(n_\psi^\mathrm{eq})^2\,, 
\ee
where $\langle\sigma v\rangle_\mathrm{fi}$ is the total cross-section for  
$\psi\psi\to\chi\chi$ and $\psi\psi\to\chi\psi$. Since $n_\psi^\mathrm{eq} \gg n_\chi$, a 
necessary condition for transformation processes 
to be non-negligible compared to traditional freeze-in
is thus $\langle\sigma v\rangle_\mathrm{fi}\ll \langle\sigma v\rangle_\mathrm{tr}$. Once the 
two terms on the r.h.s.~of the above equation are of a similar size, on the other hand, 
$\chi\psi\to\chi\chi$ will very quickly take over due to the exponential growth of $n_\chi$.

We show the evolution of the DM abundance for this scenario in Fig.~\ref{fig:epi_simp_fi}, assuming for 
simplicity that all amplitudes are constant.
We also choose the same masses and transformation 
couplings as in Fig.~\ref{fig:epi_simp}, to facilitate comparison.
Instead of fixing the initial abundance, however, we now fix the freeze-in coupling
to result in the correct relic abundance (thus taking a vanishing DM abundance as the initial condition).
The above discussed three phases -- freeze-in, 
transformation, and the final flattening of the 
abundance evolution curve -- 
are clearly visible in the figure. We stress that this brings a new perspective to the widely studied freeze-in mechanism,
which can trigger a subsequent 
phase of exponential growth in 
a rather natural way. 
It therefore becomes possible to 
satisfy the relic density constraint with significantly smaller couplings $\lambda_{\rm fi}$ than generally 
assumed, 
opening up new parameter space where freeze-in is relevant for setting the DM energy density.

\begin{figure}[t]
\includegraphics[width=0.95\columnwidth]{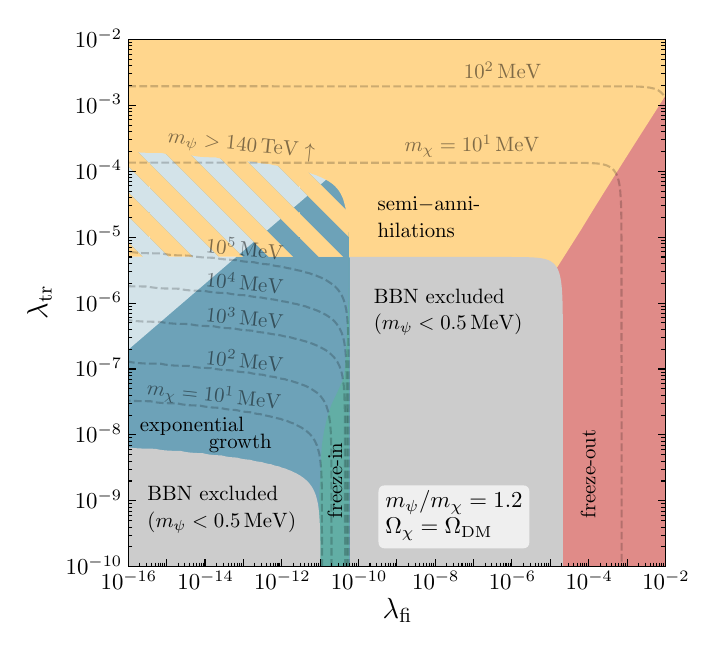}
\caption{%
{\it Phase diagram} of transformation 
($\lambda_{\rm tr}$) and freeze-in ($\lambda_{\rm fi}$) couplings that 
can result in the correct DM abundance, for a fixed mass ratio $m_\psi/m_\chi=1.2$. Colored regions 
indicate the respective mechanism that is  responsible for thermal production, while dashed lines show the
required value of $m_\chi$. Gray regions would require a new heat bath particle $\psi$ too light
to be compatible with constraints from BBN~\cite{Depta:2019lbe}. In the light blue region an additional $2\to4$ freeze-in contribution is expected.
}
\label{fig:phases}
\end{figure}

To further illustrate the last point we show in Fig.~\ref{fig:phases} a full `phase diagram' 
of the combination
of couplings $\lambda_{\rm tr}$ and $\lambda_{\rm fi}$ that allow the production of DM from the heat bath,
for a fixed mass ratio of $m_\psi/m_\chi=1.2$. At each point in this plane, we thus adjust the mass $m_\chi$ 
such that $\Omega_{\text{DM}} h^2 = 0.12$  (with dashed lines indicating isocontours of $m_\chi$).
Depending on the couplings, the relic abundance can be set via different mechanisms. In the green (red) 
region, the relic is mainly set via freeze-in (freeze-out) 
of the  process $\psi\psi \leftrightarrow \chi \chi$. 
In the blue (yellow) region, the relic density is instead mainly set via freeze-in (freeze-out) of the 
process $\psi\chi \leftrightarrow \chi\chi$. In regions where two colors overlap, more than one production mechanism can 
lead to the correct relic abundance, albeit for different masses. In the gray regions, either of the production mechanisms
would require fully thermalized scalars with $m_\psi < 0.5\,\mathrm{MeV}$ -- for the mass ratio 
$m_\psi/m_\chi=1.2$ adopted for the purpose of this figure -- which is in conflict with constraints from big 
bang nucleosynthesis (BBN)~\cite{Depta:2019lbe}. The blue region is bounded 
towards large values of 
$\lambda_\text{tr}$, 
since such couplings would -- for exponential 
production -- require masses above the unitarity limit of a 
thermal particle, $m_\psi > 140\,\mathrm{TeV}$~\cite{Smirnov:2019ngs}.

There is an irreducible  $2\to 4$ freeze-in contribution $\psi \psi \to 4 \chi$, with cross-section scaling as 
$\lambda_\mathrm{tr}^4$, 
which we estimate to dominate over $2\to 2$ freeze-in within the light blue region. 
%
%
We neglect $2\to4$ processes in Fig.~\ref{fig:phases},  which only have a logarithmic effect on the value of $\lambda_{\rm tr}$ that results in the observed relic density.
Finally, we note that $\lambda_{\rm tr}$ generates $\lambda_{\rm fi}$ radiatively and, in the absence of fine-tuning, we expect $\lambda_{\rm fi} \gtrsim \lambda_{\rm tr}^2 / (4 \pi)^2$.  This bound is satisfied except in the light blue region where $2\to4$ processes are relevant.

\smallskip
\paragraph*{Discussion.---}%
We stress that exponential growth due to processes 
as depicted in Fig.~\ref{fig:infection}
is by no means restricted to specific model 
realizations, but is a general mechanism of DM production that essentially interpolates between 
the traditionally considered freeze-in and freeze-out regimes.
At first glance it may seem worrisome that the dark matter density depends exponentially on the 
transformation 
cross-section, implying that the cross-section must be carefully chosen to match observation. But this 
can be turned around: in fact it implies that this mechanism is highly predictive, as manifested  by the 
logarithmic sensitivity of the necessary cross-section to the initial conditions.  
 We also note that exponential sensitivity is quite common in nature 
 and for example can result from
 renormalization group flows, where IR parameters can be exponentially sensitive to UV parameters: 
 the proton mass, {\it e.g.}, depends exponentially on the size of the strong coupling
 at high energies~\cite{Anderson:1994dz}.

It is worth pointing out that the mechanism proposed here works for a large range of different particle masses,
not just the specific choice displayed in the examples above.
 Limits from BBN or CMB on new light degrees of freedom~\cite{Depta:2019lbe,Sabti:2019mhn}, as encountered in Fig.~\ref{fig:phases}, could be significantly lowered by considering the possibility 
 of $\psi$ being a SM particle.
 Larger values of $m_\psi$ would not affect the DM velocity dispersion at $T\sim m_\chi$
 (but may sharpen model-dependent constraints from the decay of $\psi$, see below).
 Even a reverted mass hierarchy is possible, $m_\chi>m_\psi$, in which case it would be $m_\chi$ 
rather than $m_\psi$ that determines when the evolution of $n_\chi/s$ starts to flatten.
For  $m_\chi>m_\psi$, DM in general becomes unstable -- but not necessarily on cosmological
time-scales.
 Such highly suppressed decays may potentially be visible in late-time observables related to 
  cosmological structure formation or cosmic rays, making corresponding scenarios even more attractive. 
 
 \begin{figure}[t]
\includegraphics[width=0.95\columnwidth]{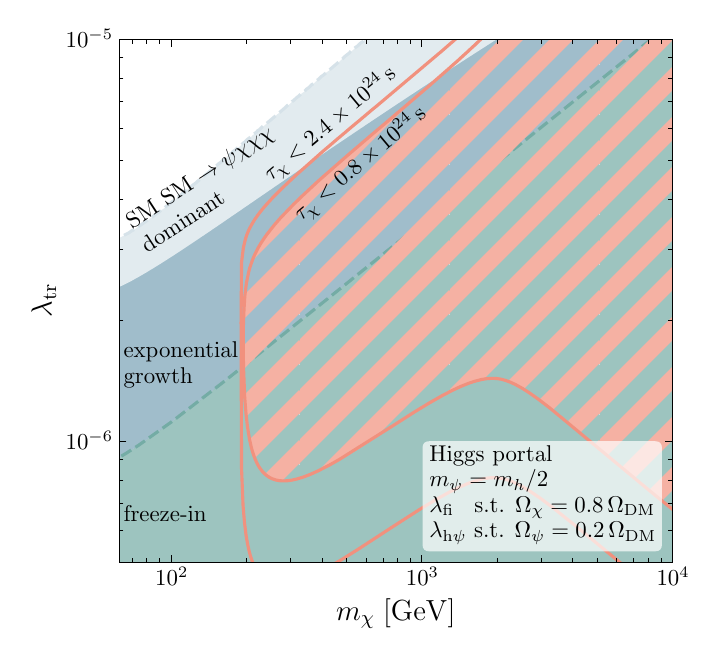}
\caption{%
Phenomenological consequences of a specific example realization of exponential production, where $\psi$ is 
coupled to the SM via the Higgs portal in the regime of resonant annihilations ($m_\psi = m_h / 2$). We fix 
$\lambda_{h \psi} = 3.9 \times 10^{-4}$, giving $20 \%$ of DM in $\psi$, and choose $\lambda_\mathrm{fi}$ for 
each $m_\chi$ and $\lambda_\mathrm{tr}$ such that $\chi$ provides $80 \%$ of the observed DM abundance. 
Green and blue regions indicate which production mechanism dominates the $\chi$-abundance. Red lines (and 
hatched area) show where CMB signatures can naturally be expected. Within the light blue region, freeze-in via
$\mathrm{SM \, SM} \to \psi \chi \chi \chi$ dominates over $\chi \chi \to \psi \psi$ in setting the initial $\chi$ abundance, thus generically 
leading to overproduction of DM.
}
\label{fig:pheno}
\end{figure}

Let us finish by considering the phenomenological consequences for a simple, concrete model realization,
where $\psi$ is a real scalar and couples to the SM via a quartic Higgs portal coupling 
$\lambda_{h \psi} |H|^2 \psi^2 / 2$.
The relic density of $\psi$ is then determined by standard freeze-out, exactly as for scalar singlet 
DM~\cite{Cline:2013gha} where the resonant region $m_\psi\sim m_h/2$ features the best likelihood
in globals fits~\cite{GAMBIT:2017gge}. For concreteness we use $m_\psi= m_h/2$
and  $\lambda_{h \psi}=3.9 \times 10^{-4}$, resulting in $\Omega_\psi = 0.2 \, \Omega_\mathrm{DM}$ when
including the effect of early kinetic decoupling as implemented in DarkSUSY~\cite{Binder:2021bmg}.  
In Fig.~\ref{fig:pheno}, we show the plane $m_\chi - \lambda_\mathrm{tr}$ after fixing 
$\lambda_\mathrm{fi}$  such that the dominant DM component satisfies 
$\Omega_\chi = 0.8 \, \Omega_\mathrm{DM}$ {\it and} is mostly produced by transformation 
(blue) or $\psi\psi\to\chi\chi$ freeze-in  (green) processes; in the white area 
$\Omega_\chi = 0.8 \, \Omega_\mathrm{DM}$ may only be possible via freeze-out or semi-annihilations. 
Using \textsc{FeynRules}~\cite{Alloul:2013bka} and \textsc{MadGraph}5~\cite{Alwall:2011uj}, furthermore,
we find that for $\lambda_\mathrm{fi} \lesssim \lambda_{h \psi} \lambda_\mathrm{tr} / [3 (4 \pi)^2]$
freeze-in production through 
$\mathrm{SM} \, \mathrm{SM} \to \psi \chi \chi \chi$ would dominate over that from $\psi\psi\to\chi\chi$
(indicated by the light blue region).
Finally, the decay of $\chi$ to SM particles can impact the observed CMB power spectrum; 
for an 80\% DM subcomponent, this leads to a  constraint 
of $\tau_\chi \gtrsim 0.8\times10^{24} \, \mathrm{s}$~\cite{Slatyer:2016qyl}
(indicated by the hatched area in Fig.~\ref{fig:pheno}), projected to tighten by a factor 
of $\sim$3 with CMB-S4~\cite{Abazajian:2016yjj} (red lines). 
Here we estimated the decay width as
\begin{align}
\Gamma_\chi \simeq &\frac{\lambda_{h \psi}^2 v^2}{16 \pi m_\chi^3} \frac{\lambda_\mathrm{tr}^2 \lambda_\mathrm{fi}^2}{(4 \pi)^8} \int_0^{(m_\chi - m_\psi)^2} \frac{\D q^2}{2 \pi}  2 q\, \Gamma_h (q) \nonumber \\
&\times \frac{\sqrt{[m_\chi^2 - (m_\psi + q)^2][m_\chi^2 - (m_\psi - q)^2]}}{(q^2 - m_h^2)^2 + m_h^2 \Gamma_h^2 (m_h)}\,, \label{eq:Gamma_x}
\end{align}
where $v \approx 246 \, \mathrm{GeV}$ is the vacuum expectation value of the Higgs, and $\Gamma_h(q)$ is 
its off-shell decay width as evaluated with HDECAY~\cite{Djouadi:2018xqq}. 
For $m_\chi < m_\psi + m_h = (3/2) m_h$, the Higgs can only be produced off-shell, resulting in a
significant suppression of $\Gamma_\chi$ and hence the absence of CMB limits in this parameter range.

\smallskip
\paragraph*{Conclusions.---}%
We have introduced a novel type of DM production mechanism,
where an initially tiny DM abundance 
is enhanced due to a 
process where DM particles convert bath particles into more DM particles.  The DM 
abundance grows exponentially with time, 
in stark contrast to 
models of freeze-in where the abundance grows only as a power law.  
Our mechanism complements both freeze-in and freeze-out thermal production 
scenarios in a generic way. Concrete model realizations lead, already in their simplest forms, to interesting 
phenomenological consequences.
Further, and detailed, exploration of this new 
way of producing DM from the thermal bath
thus appears highly warranted.

\vspace*{2cm}
\noindent
{\it Note added.}
After our work appeared on the arXiv, the mechanism introduced here was independently
suggested in Ref.~\cite{Hryczuk:2021qtz}. In that reference, the authors study in detail the 
phenomenological consequences of another concrete model realization of Pandemic DM.

\vfill
 \paragraph*{Acknowledgements.---}%
We thank Michael Geller for helpful conversations, including at a primordial stage.  JTR thanks the participants of the workshop ``New Physics from the Sky" at the Galileo Galilei Institute for helpful feedback. This work is supported by the Deutsche Forschungsgemeinschaft under Germany’s
Excellence Strategy – EXC 2121 ‘Quantum Universe’ – 390833306, the F.R.S. – FNRS under the Excellence of Science 
(EoS) project No. 30820817 – be.h ‘The H boson gateway to physics beyond the Standard Model,’ and the National 
Science Foundation under Grant No. NSF PHY-1748958.  JTR is further supported by the NSF CAREER grant 
PHY-1554858, NSF grant PHY-1915409, an award from the Alexander von Humboldt Foundation, and the European Research Council (ERC) under the EU Horizon 2020 Programme (ERC-CoG-2015 - Proposal n. 682676 LDMThExp).
\newpage

\bibliography{pandemic.bib}

\end{document}